\documentclass[aps,prc,
superscriptaddress,
preprint,
showpacs]{revtex4-1}
\usepackage{amsmath}
\usepackage{color}
\usepackage{graphicx}
\usepackage{float}
\begin{document}
\title{Investigation of the $^{9}$B nucleus and its cluster-nucleon correlations}
\author{Qing Zhao} \email{zhaoqing91@outlook.com.}
\affiliation{School of Physics and Key Laboratory of Modern Acoustics,
Institute of Acoustics, Nanjing University, Nanjing 210093, China}

\author{Zhongzhou Ren} \email{zren@nju.edu.cn, zren@tongji.edu.cn.}  \affiliation{School
  of Physics and Key Laboratory of Modern Acoustics, Institute of
  Acoustics, Nanjing University, Nanjing 210093, China}
\affiliation{School of Physics Science and Engineering,
Tongji University, Shanghai 200092, China}

\author{Mengjiao Lyu} \email{mengjiao@rcnp.osaka-u.ac.jp}
\affiliation{School of Physics and Key Laboratory of Modern Acoustics,
Institute of Acoustics, Nanjing University, Nanjing 210093, China}
\affiliation{Research Center for Nuclear Physics (RCNP), Osaka
  University, Osaka 567-0047, Japan}

\author{Hisashi Horiuchi} \affiliation {Research Center for Nuclear
  Physics (RCNP), Osaka University, Osaka 567-0047, Japan}
\affiliation {International Institute for Advanced Studies, Kizugawa
  619-0225, Japan}

\author{Yasuro Funaki} \affiliation{Laboratory of Physics, Kanto
  Gakuin University, Yokohama 236-8501, Japan}

\author{\mbox{Gerd R\"{o}pke}} \affiliation{Institut f\"{u}r Physik,
  Universit\"{a}t Rostock, D-18051 Rostock, Germany}

\author{Peter Schuck} \affiliation{Institut de Physique Nucl\'{e}aire,
  Universit\'e Paris-Sud, IN2P3-CNRS, UMR 8608, F-91406, Orsay,
  France} \affiliation{Laboratoire de Physique et Mod\'elisation des
  Milieux Condens\'es, CNRS-UMR 5493, F-38042 Grenoble Cedex 9,
  France}

\author{Akihiro Tohsaki} \affiliation{Research Center for Nuclear
  Physics (RCNP), Osaka University, Osaka 567-0047, Japan}

\author{Chang Xu} \affiliation{School of Physics and Key Laboratory of
  Modern Acoustics, Institute of Acoustics, Nanjing University,
  Nanjing 210093, China}

\author{Taiichi Yamada} \affiliation{Laboratory of Physics, Kanto
  Gakuin University, Yokohama 236-8501, Japan}

\author{Bo Zhou} \affiliation{Meme Media Laboratory, Hokkaido University, Sapporo
  060-8628, Japan}

\begin{abstract}
  In order to study the correlation between clusters and nucleons in light nuclei, we formulate a new superposed THSR wave function which describes both spatial large spreading and cluster-correlated dynamics of valence nucleons. By using the new THSR wave function, the binding energy of $^{9}$B is essentially improved comparing with our previous studies. We calculate the excited states of $^{9}$B and obtain the energy spectrum of $^9$B which is consistent with the experimental results, including prediction of the $1/2^+$ excited state of $^9$B which is not fixed yet experimentally. We study the proton dynamics in $^{9}$B and find that the cluster-proton correlation plays an essential role for the proton dynamics in the ground state of $^{9}$B. Further more, we discuss the density distribution of the valence proton with special attention to its tail structure. Finally, the resonance nature of excited states of $^9$B is illustrated by comparing root-mean-square radii between the ground and excited states.
\end{abstract}

\maketitle

\section{Introduction}
As one of the most important effects in nuclear physics, the clustering degree of freedom has been extensively studied for various nuclear many-body problems \cite{Yamada2005, Funaki2015, Freer2017, Tohsaki2001, Funaki2002, Zhou2012, Suhara2014, Zhou2013, Zhou2014, Lyu2015, En'yo1995, Xu2006, Ren2012, He2014, Yang2014}, including the microscopic description of cluster states \cite{Zhou2013, Zhou2012} and cluster radioactivity such as the $\alpha$ decay process \cite{Xu2006, Ren2012}. With the discovery of $\alpha$-cluster condensation in light nuclei, the Tohsaki-Horiuchi-Schuck-R\"{o}pke(THSR) wave function is proposed to describe gas-like cluster states in light nuclei, especially for $^{12}$C and $^{16}$O \cite{Tohsaki2001, Funaki2002, Zhou2012, Zhou2013, Yamada2005, Suhara2014, Zhou2014, Lyu2015}. Investigations with the THSR wave function prove the nonlocalised property of cluster dynamics in light nuclei such as $^{20}$Ne \cite{Zhou2013}. By comparing with generator coordinate method (GCM), which can be regarded as an advanced solution of the nuclear many-body system, clustering states are found to be almost 100\% accurately described by a single THSR wave function \cite{Funaki2003, Funaki2005, Funaki2009, Zhou2013, Zhou2012}.

For the study of non-conjugate nuclei composed of both $\alpha$-clusters and valence nucleons, the THSR wave function has been successfully extended to the $^{9-11}$Be isotopes \cite{Lyu2015,Lyu2016,Lyu2017}. It is found that the cluster-cluster correlation and nucleon-nucleon correlation can be well described by the THSR wave function in these systems \cite{Lyu2016, Zhao10B}. However, there is few studies devoted to the correlation between  $\alpha$-clusters and valence nucleons. In order to study dynamics of clusters and valence nucleons under the cluster-nucleon correlation, we superpose the THSR wave function to describe both large spreading and cluster-correlated configurations for valence nucleons in light nuclei. Here, we choose the nucleus $^{9}$B, which has $\alpha+\alpha+p$ cluster structure, as the first application of our new framework. Due to the lack of good present understanding of the  $^{9}$B nucleus, it is also desirable to further study this proton-rich nucleus with the new microscopic model.

In Section \ref{sec:waveFunction} we formulate the superposed THSR wave function for ${}^{9}$B. Then in Section \ref{sec:results} we show the results for the physical properties of ${}^{9}$B and discuss the structure and dynamics of the clustering states, especially the cluster-nucleon correlation effect. The last Section \ref{sec:conclusion} contains the conclusions.

\section{Formulation of THSR Wave Function of ${}^{9}$B}
\label{sec:waveFunction}

We first write the THSR wave function for $^{9}$B with the form used in our previous works \cite{Lyu2015}, as
\begin{equation}\label{eq:old-thsr}
   \begin{split}
\Phi = \prod_{i=1}^2\int &d\mathbf{R}_i{\rm exp}(-\frac{R_{i,x}^2}{\beta_{\alpha,xy}^2}-\frac{R_{i,y}^2}{\beta_{\alpha,xy}^2}-\frac{R_{i,z}^2}{\beta_{\alpha,z}^2})\\
&\times\int d\mathbf{R}_p{\rm exp}(-\frac{R_{p,x}^2}{\beta^2_{p,xy}}-\frac{R_{p,y}^2}{\beta^2_{p,xy}}-\frac{R_{p,z}^2}{\beta^2_{p,z}})\\
&\times
e^{im\phi_{\mathbf{R}_{p}}}\Phi^B(\mathbf{R}_{1},\mathbf{R}_{2},\mathbf{R}_{p}),
   \end{split}
\end{equation}
where $\beta_{\alpha}$s are Gaussian parameters for the
nonlocalized motion of two $\alpha$-clusters and $\beta_{p}$s
are Gaussian parameters for valence proton.  $e^{im\phi_{\mathbf{R}_{p}}}$ is the phase factor which determines the intrinsic parity of the wave function \cite{Lyu2015}. Parameter $m=1$ corresponds to the intrinsic negative parity while $m=0$ corresponds to the intrinsic positive parity.
$\Phi^B$ is the Brink wave function which is writen as
\begin{equation}
   \begin{split}
\Phi^B(R_1, R_2, R_p) = \mathcal{A}\{\psi(\alpha_1, R_1)\psi(\alpha_2, R_2)\psi(r_p, R_p)\},
   \end{split}
\end{equation}
where $R_{(1,2)}$ and $R_p$ are corresponding generator coordinates for the $\alpha$ clusters and valence proton, respectively.

To investigate correlation between the extra proton and each $\alpha$-cluster, we formulate the THSR wave function of ${}^{9}$B as a superposition of both cluster-correlated configuration and large spreading configuration which is named by the "superposed THSR wave function", as
\begin{equation}
\Psi = c\Phi_1+c\Phi_2+d\Phi_{3}.
\end{equation}
Here $\Phi_1$ and $\Phi_2$ are cluster-correlated configurations of $^9$B which describes the correlated motion of valence proton around each $\alpha$-cluster. $\Phi_{3}$ is a term corresponding to the large spreading configuration of $^9$B in which the valence proton orbits around the $^8$Be core. Here parameters $c$ and $d$ are coefficient parameters of the superposed THSR wave function.

The THSR wave function in Eq.~\ref{eq:old-thsr}, which is named by "traditional THSR wave function", is used as component $\Phi_{3}$. The terms corresponding to the correlation configuration $\Phi_1$ and $\Phi_2$ are constructed by replacing the generator coordinate $R_p$ of the valence proton in $\Phi_{3}$ with coordinate $R_p + R_j$, as
\begin{equation}
   \begin{split}
\Phi_j(j = 1, 2) = \prod_{i=1}^2\int &d\mathbf{R}_i{\rm exp}(-\frac{R_{i,x}^2}{\beta_{\alpha,xy}^2}-\frac{R_{i,y}^2}{\beta_{\alpha,xy}^2}-\frac{R_{i,z}^2}{\beta_{\alpha,z}^2})\\
&\times\int d\mathbf{R}_p{\rm exp}(-\frac{R_{p,x}^2}{\beta'^2_{p,xy}}-\frac{R_{p,y}^2}{\beta'^2_{p,xy}}-\frac{R_{p,z}^2}{\beta'^2_{p,z}})\\
&\times
e^{im\phi_{\mathbf{R}_{p}+\mathbf{R}_{j}}}\Phi^B(\mathbf{R}_{1},\mathbf{R}_{2},\mathbf{R}_{p}+\mathbf{R}_{j}).
   \end{split}
\end{equation}
Here, the generator coordinate $R_{p}$ is used to describe the correlated motion of the valence proton around the $\alpha$-cluster with generator coordinate $R_j$, whose subscript $j$ denotes each of the two $\alpha$-clusters.

We also apply the angular-momentum projection technique
$\hat{P}_{MK}^{J}\left| \Psi \right\rangle$ to restore the rotational symmetry \cite{Schuck1980},
\begin{equation}
  \begin{split}
\left| \Psi^{JM} \right\rangle&=\hat{P}_{MK}^{J}\left| \Psi \right\rangle\\
    &=\frac{2J+1}{8\pi^{2}}\int d \Omega D^{J*}_{MK}(\Omega)\hat R (\Omega)
    \left| \Psi \right\rangle,
  \end{split}
\end{equation}
where $J$ is the total angular momentum of ${}^9$B.

The Hamiltonian of the ${}^{9}$B nucleus can be written as
\begin{equation}\label{hamiltonian}
  H=\sum_{i=1}^{9} T_i-T_{c.m.} +\sum_{i<j}^{9}V^N_{ij}
    +\sum_{i<j}^{9}V^C_{ij} +\sum_{i<j}^{9}V^{ls}_{ij}.
\end{equation}
Here $T_{c.m.}$ is the spurious kinetic energy of the center-of-mass
motion. Volkov No. 2 \cite{Volkov1965} interaction is selected as the central force of the nucleon-nucleon potential as
\begin{equation}\label{vn}
  V^N_{ij}=\{V_1 e^{-\alpha_1 r^2_{ij}}-V_2 e^{-\alpha_2 r^2_{ij}}\}
  \{ W - M \hat P_\sigma \hat P_\tau \ + B \hat P_{\sigma} - H \hat P_{\tau}\},
\end{equation}
where $M=0.6$, $W=0.4$, $B=H=0.125$, $V_{1}=-60.650$ MeV, $V_{2}=61.140$ MeV,  $\alpha_{1}=0.309 $fm${}^{-2}$, and $\alpha_{2}=0.980$ fm${}^{-2}$.

The Gaussian soft core potential with three ranges (G3RS) term \cite{Yamaguchi1979} is taken as the spin-orbit interaction,
\begin{equation}\label{vc}
V^{ls}_{ij}=V^{ls}_0\{
           e^{-\alpha_1 r^2_{ij}}-e^{-\alpha_2 r^2_{ij}}
           \} \mathbf{L}\cdot\mathbf{S} \hat{P}_{31},
\end{equation}
where $\hat P_{31}$ projects the two-body system into triplet odd
state.  Parameters in $V^{ls}_{ij}$ are taken as $V_{0}^{ls}$=2000 MeV, $\alpha_{1}$=5.00 fm${}^{-2}$, and $\alpha_{2}$=2.778 fm${}^{-2}$ from Ref. \cite{Okabe1979}.

The Gaussian size parameter $b$ for single-nucleon states is set to be $b=1.35$ fm, which is the same as that used in our previous work in Ref.~\cite{Lyu2015}.

\section{Results and Discussions}
\label{sec:results}

We discuss the energy spectrum of the $^9$B nucleus obtained with the new superposed THSR wave function $\Psi$ and compare them with those obtained by the tradional THSR wave function ($\Phi_{3}$) used in our previous work \cite{Lyu2015}, as shown in Table.~\ref{table:result2}. Both of these THSR wave functions $\Phi$ and $\Phi_3$ are variationally optimized. In this table, we also include the GCM results for the spectrum of $^9$B, to justify the accuracy of  descriptions by the THSR wave functions.
\begin{table*}[htbp]
  \begin{center}
    \caption{\label{table:result2}The $3/2^{-}$ rotational band of ${}^{9}$B. $E^{\text{THSR}}_{old}$ denotes results obtained with the traditional THSR wave function $\Phi_{3}$ used in our previous work \cite{Lyu2015}. $E^{\text{THSR}}_{new}$ denotes results obtained with our new superposed THSR wave function $\Psi$. $E^{\text{GCM}}$ denotes results obtained with GCM calculation. Values in parentheses are corresponding excitation energies. $\Delta$ denotes the improvement obtained by our new superposed THSR wave function.  All units of energies are in MeV.}
 \begin{tabular*}{12cm}{ @{\extracolsep{\fill}} l c c c c c}
    \hline
    \hline
State &$E^{\text{THSR}}_{old}$ &$E^{\text{THSR}}_{new}$ &$\Delta$ &$E^{\text{GCM}}$ &$E^{\text{Exp}}$ \cite{Tilley2004, Audi2012}\\
    \hline
$7/2^-$(E.S.) &-47.0(6.9) &-48.3 (6.6) &1.3 &-48.1(7.1) &-49.3 (7.0)\\
$5/2^-$(E.S.) &-51.4(2.5) &-52.4 (2.5) &1.0 &-52.9(2.3) &-53.9 (2.4)\\
$3/2^-$(G.S.) &-53.9      &-54.9       &1.0 &-55.2      &-56.3      \\
    \hline
    \hline
  \end{tabular*}
  \end{center}
\end{table*}

\begin{figure}[htbp]
  \centering
  \includegraphics[width=0.6\textwidth]{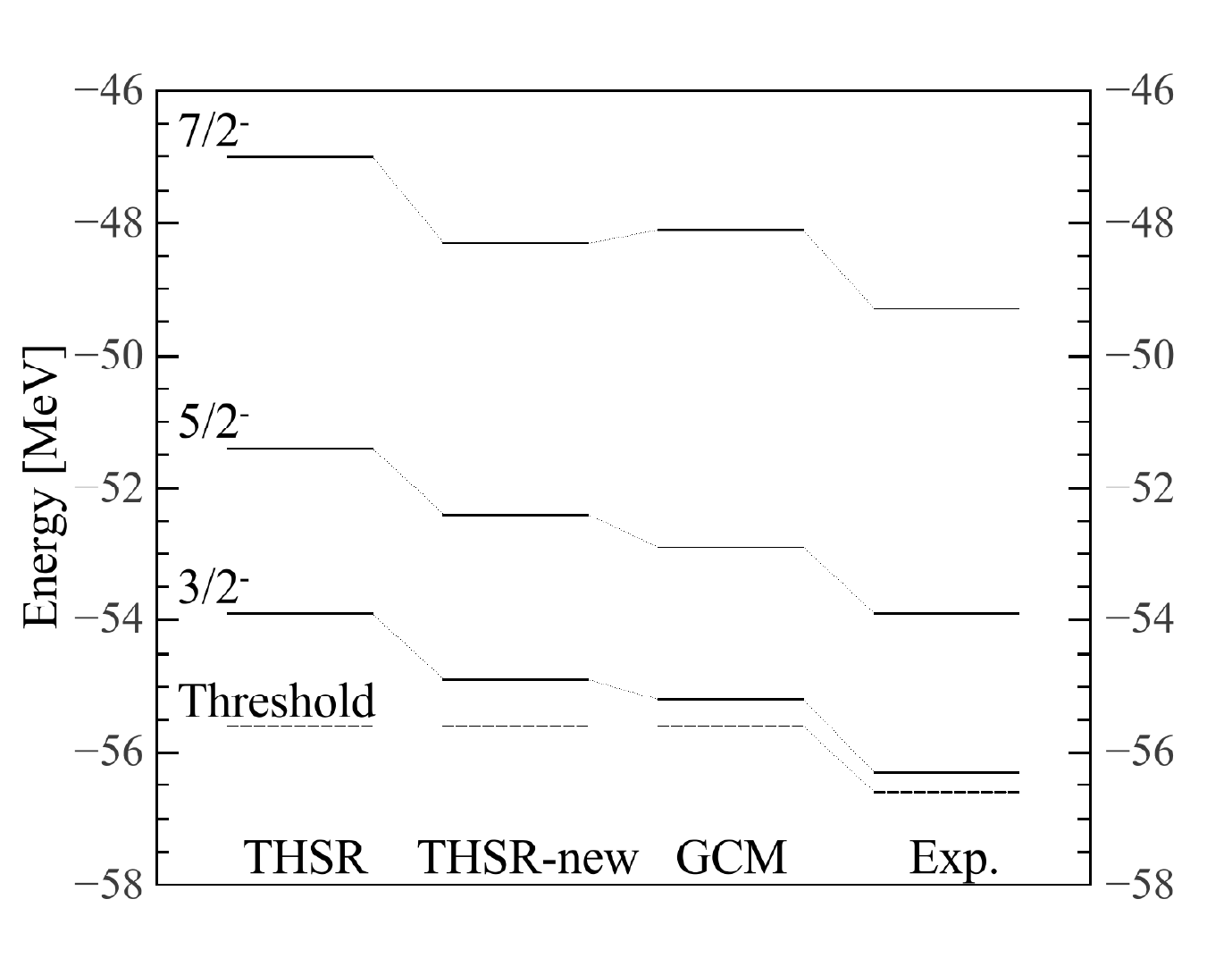}
  \caption{\label{fig:enelevel}Theoretical and experimental results of the energy spectrum of ${}^{9}$B. }
\end{figure}

From Table.~\ref{table:result2} we can observe significant improvements for the binding energy of $^9$B using our new superposed THSR wave function which are in much better agreements with the experimental values. Further more, we get almost the same results comparing with the results from GCM calculation in which 81 Brink basis states are superposed which is a strong justification of our new wave function. The ground state rotational band of ${}^{9}$B from theoretical calculations and experiments are presented in Fig.~\ref{fig:enelevel}. The dashed lines denote the corresponding $\alpha+\alpha+p$ threshold. Comparing with the traditional THSR wave function, the new superposed THSR wave function produces lower and denser energy levels  which is consistent with GCM and experiment results.

\begin{figure}[htbp]
  \centering
  \includegraphics[width=0.6\textwidth]{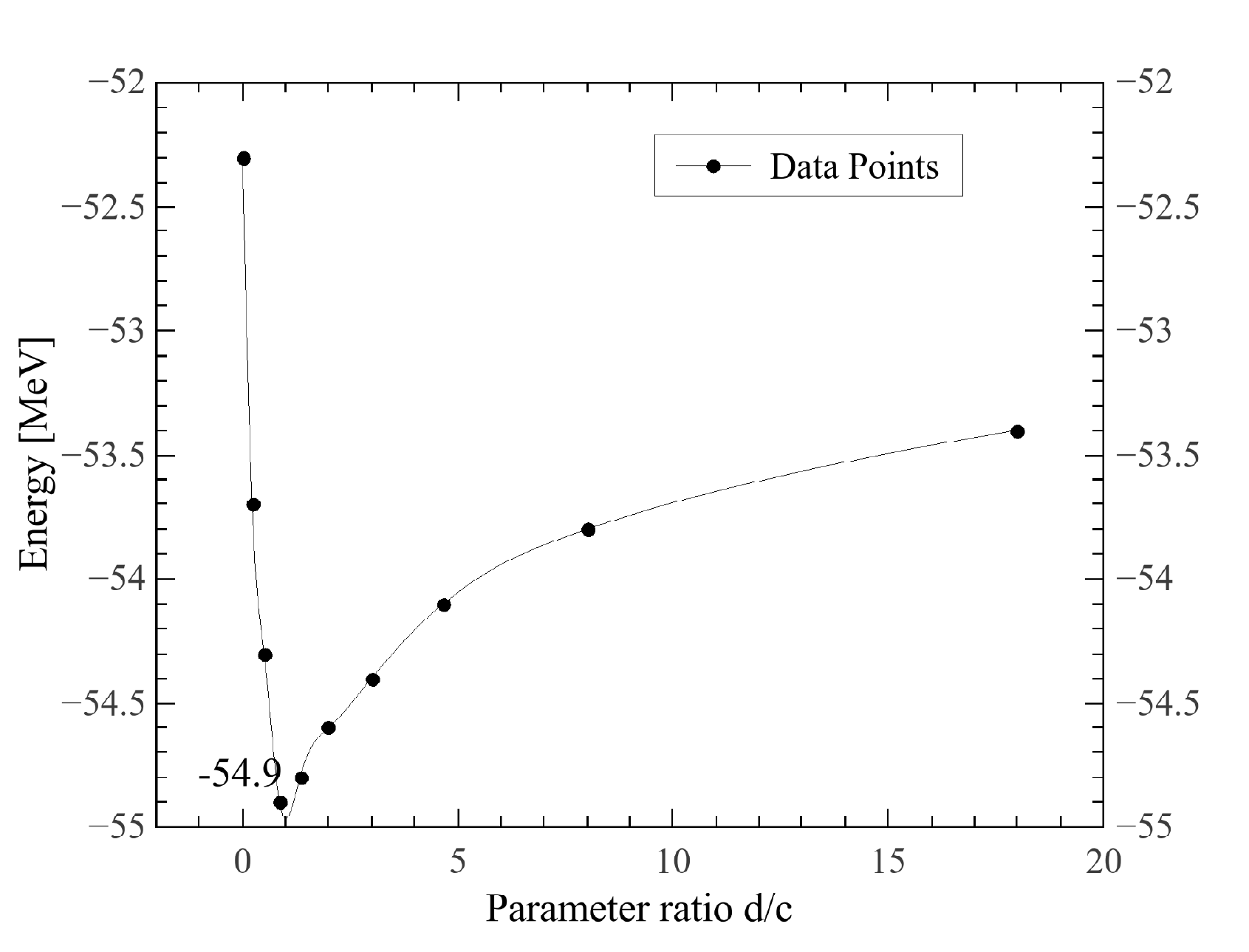}
  \caption{\label{fig:enecurve}Energy curve of the ground state of $^9$B with respect to the ratio $d/c$, while other parameters are fixed as the same. }
\end{figure}

By analyzing the analytical form of the new superposed THSR wave function $\Psi = c\Phi_1+c\Phi_2+d\Phi_{3}$, we see that when parameter $c=0$, the total wave function $\Psi$ reduces to the traditional THSR wave function $\Phi_{3}$, which corresponds to a large spreading motion of extra proton around the $^8$Be core. When $d=0$, the total wave function $\Psi$ reduces to $\Phi_{1}+\Phi_{2}$, which describes strong correlation between the extra proton and single $\alpha$-cluster. Since parameters $c$ and $d$  are not independent with each other, we optimize the ratio $d/c$ by the variational calculation, which corresponds to the coupling of these two different configurations. We show in Table \ref{table:parameter} the variationally optimized parameters and corresponding energy for the ground state of $^9$B comparing with the other two extreme cases. The variational optimized coefficients are shown in the second row of Table \ref{table:parameter} as $c=0.34$ and $d=0.32$, with corresponding ratio $d/c=0.94$. It is shown that the ground state energy of $^9$B with optimized parameters is much lower than in the  other two extreme cases. From the variational principle, we conclude that the coupling of two different configurations is energetically preferable and the new superposed THSR wave function provides a much better description than the traditional THSR wave function which corresponds to the $c=0$ extreme case. We also show the energy curve of the ground state of $^9$B with respect to ratio $d/c$ in Fig.~\ref{fig:enecurve}, in which the minimum point of variational calculation can be seen clearly. Due to the nearly equal values of parameter $c$ and $d$, we can conclude that the cluster-correlated and the large spreading configurations both have significant contributions to the dynamics of the ground state of $^9$B. The difference in values of the parameter $\beta_{\alpha,z}$ between traditional and the new THSR wave function, shows that the distances between two $\alpha$-clusters are also different in this two descriptions, which tends to be larger in spatial spreading than when the cluster-nucleon correlation is taken into account.

\begin{table*}[htbp]
  \begin{center}
    \caption{\label{table:parameter}Variationally optimized parameters and results for the ground state of $^9$B. The first line denotes
      results with fixed $c=0$, the second line denotes
      results with variationally optimized $c=0.34$ and $d=0.32$, and the last
      line denotes results with fixed $d=0$. The remaining
	parameters are fixed to be the same. Units
      of $\beta$ are in fm, and units of energies are in MeV.}
 \begin{tabular*}{13cm}{ @{\extracolsep{\fill}} c c c c c c c c c}
    \hline
    \hline
$c$ &$d$ &$\beta_{\alpha,xy}$ &$\beta_{\alpha,z}$ &$\beta_{p,xy}$ &$\beta_{p,z}$ &$\beta'_{p,xy}$ &$\beta'_{p,z}$ &$E$\\
    \hline
0    &1.0  &0.1 &4.2  &2.5 &2.6 &/   &/   &-53.9\\
0.34 &0.32 &0.1 &5.5  &5.9 &3.1 &0.5 &2.5 &-54.9\\
0.5  &0    &0.1 &5.4  &/   &/   &0.7 &2.6 &-52.3\\
    \hline
    \hline
  \end{tabular*}
  \end{center}
\end{table*}

\begin{figure}[htbp]
  \centering
  \includegraphics[width=0.45\textwidth]{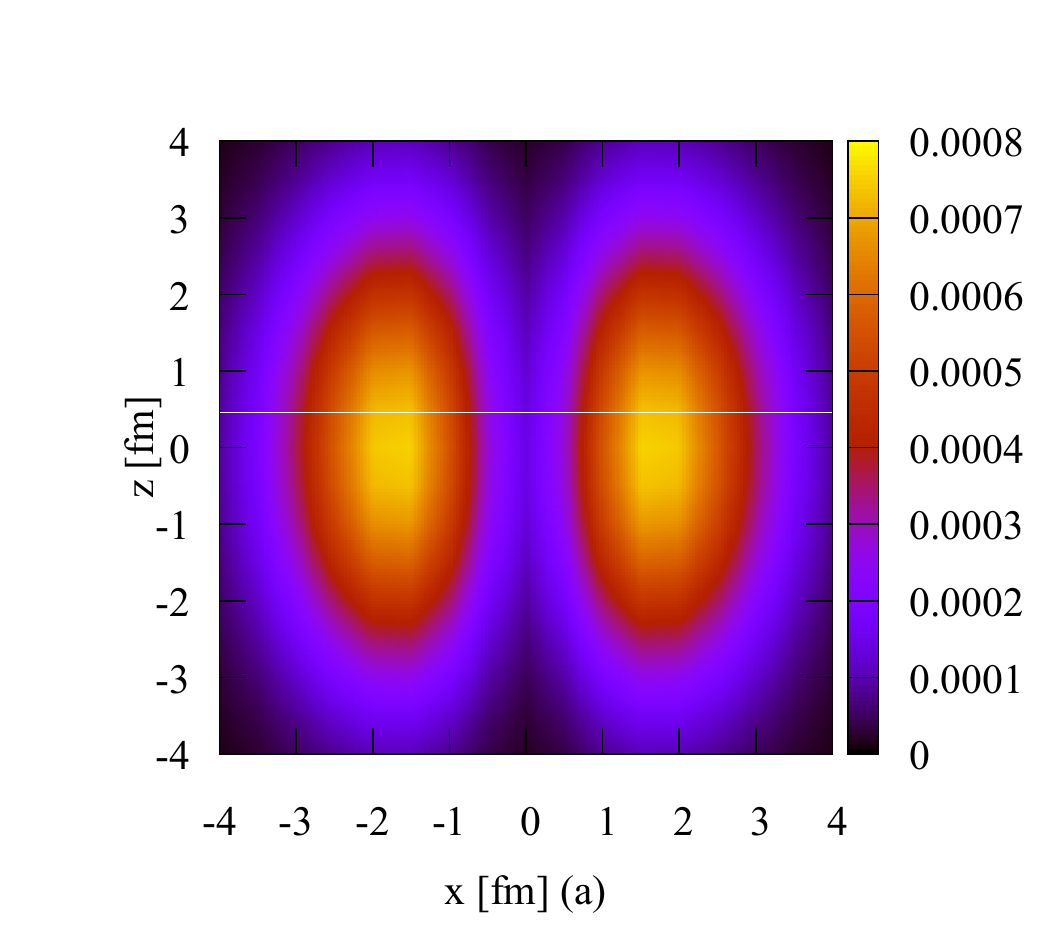}
  \includegraphics[width=0.45\textwidth]{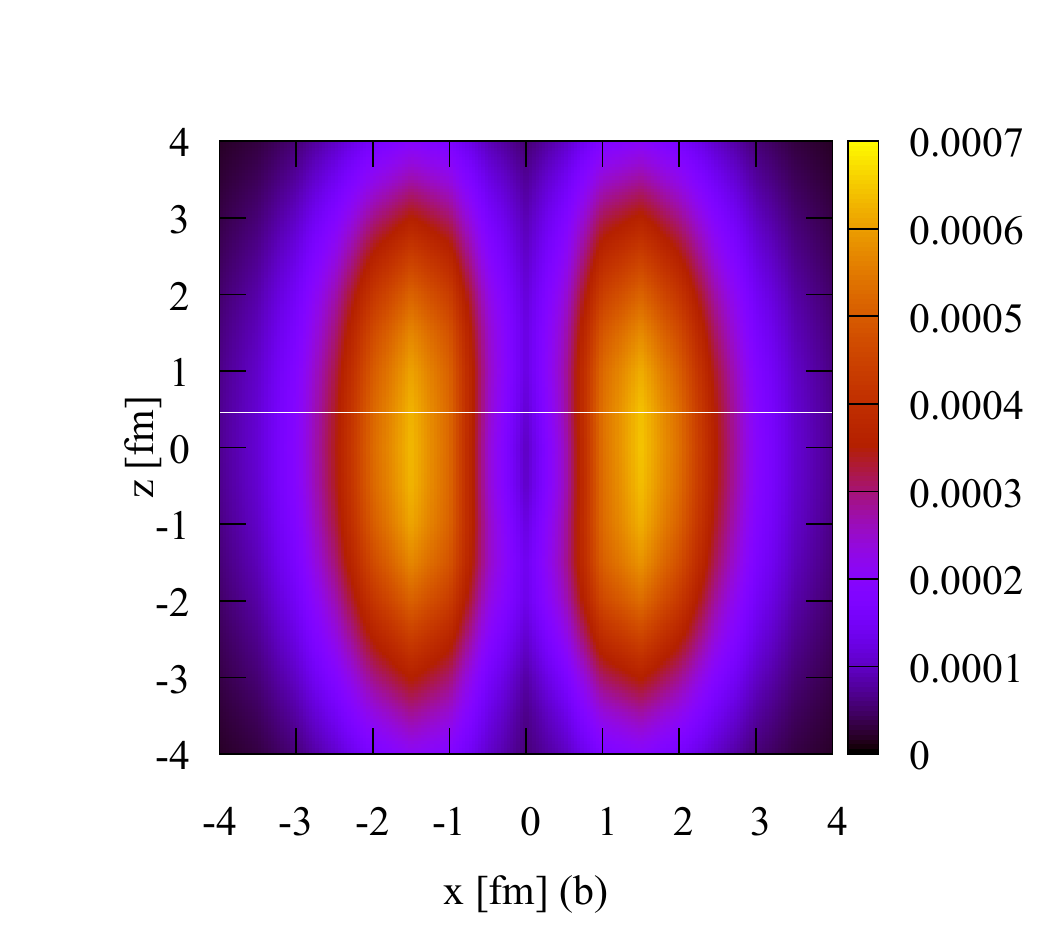}
  \caption{\label{fig:densitycompare}Density distributions of the extra proton of $^9$B obtained from different THSR wave functions. The left figure corresponds to traditional THSR wave function. The right one corresponds to the new THSR wave function.}
\end{figure}

We compare the density distributions of the valence protons obtained from the traditional and the new superposed THSR wave function as shown in Fig.~\ref{fig:densitycompare}. It is clearly observed that the density distribution of extra proton described by the new THSR wave function, is more compact in the x- and y-directions comparing with the traditional one. This is because of the fact that in the new THSR wave function, correlation effects between the extra proton and $\alpha-$clusters which are located near the z-axis in the intrinsic frame are taken into account.

\begin{figure}[htbp]
  \centering
  \includegraphics[width=0.45\textwidth]{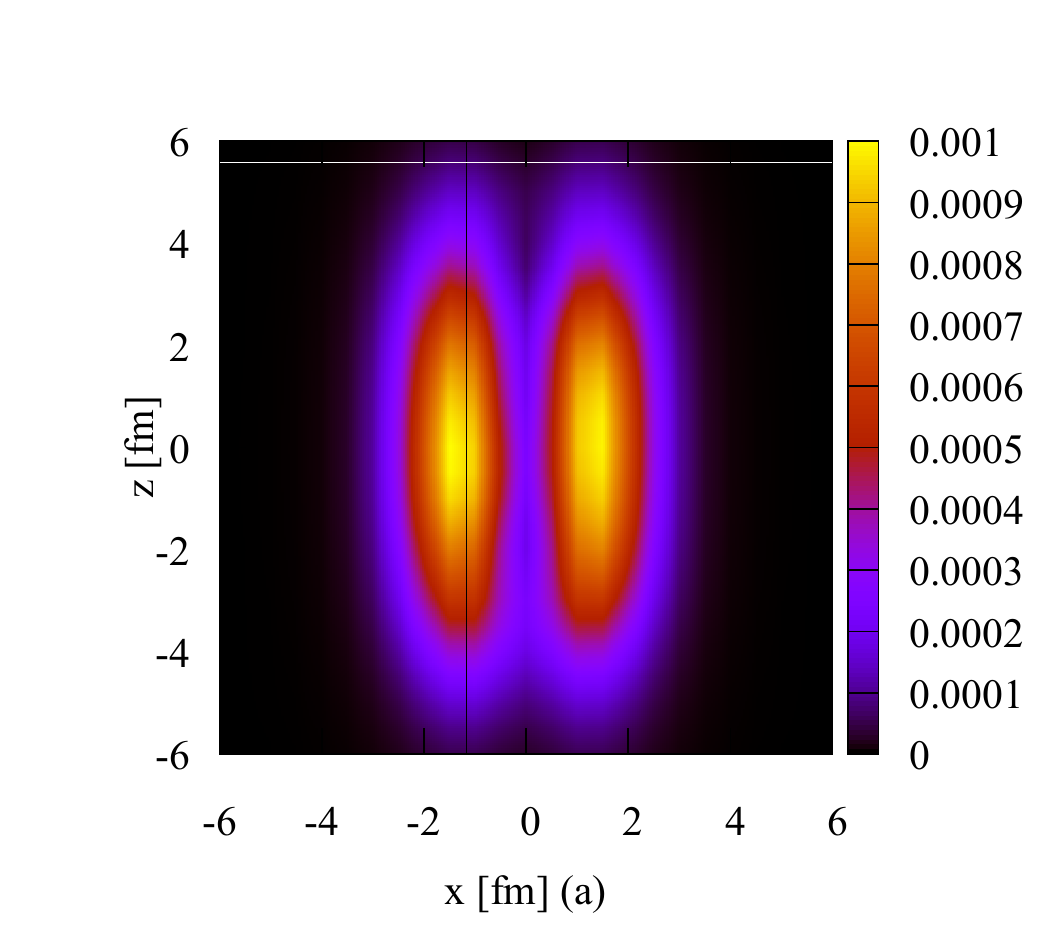}
  \includegraphics[width=0.45\textwidth]{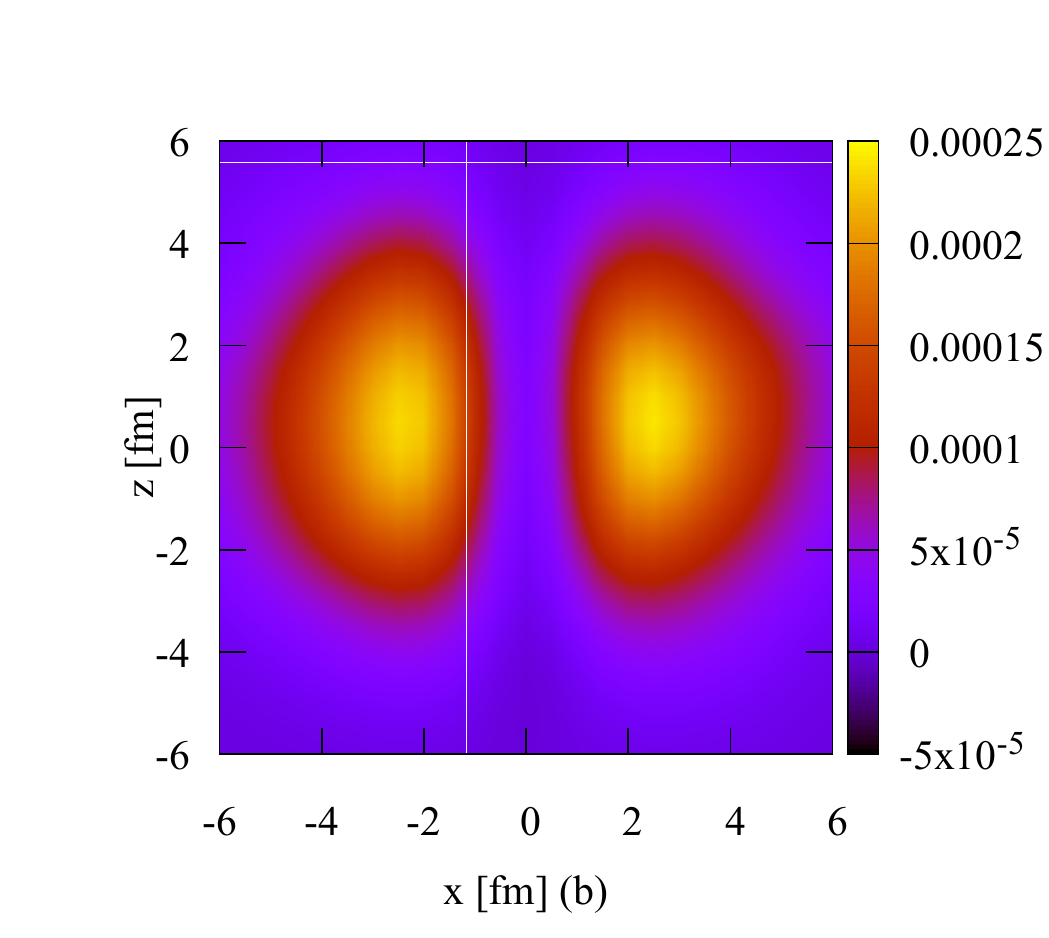}
  \caption{\label{fig:densityExtra}Density distributions of the
    extra proton in the $y=0$ cross section obtained from different terms in the new superposed THSR wave function. The left corresponds to the cluster-correlated terms $\Phi_1+\Phi_{2}$. The right corresponds to the large spreading configuration $\Phi_{3}$. Parameter $\beta$s are selected as the optimized values of the new superposed THSR wave function.}
\end{figure}

 To provide an explicit picture for the proton dynamics in $^{9}$B, we also show the density distributions of the extra proton for separate term $\Phi_1+\Phi_2$ and $\Phi_3$ of the total wave function $\Psi$, as shown in Fig.~\ref{fig:densityExtra}. We can see that the valence proton in the cluster-correlated configuration described by $\Phi_1+\Phi_{2}$ have a similar compact distribution as shown in the left panel (a), while right panel (b) shows the remaining weak correlation effect between the $^8$Be core and the valence nucleon as suggested by the much larger spread over the space. We conclude that the inner region, which is close to the $\alpha$-clusters, is mostly contributed by correlated configuration of valence proton in $\Phi_1+\Phi_2$, while the outer region is mostly contributed by the large spreading configuration of valence proton described by $\Psi_3$. It should be noticed that, even though the extra proton is strongly correlated to each $\alpha$-clusters, this does not mean that the motion of the extra proton is localized, because the $\alpha$-clusters themselves are  performing nonlocalized motion as discussed in Ref.~\cite{Lyu2016}.

\begin{figure}[htbp]
  \centering
  \includegraphics[width=0.6\textwidth]{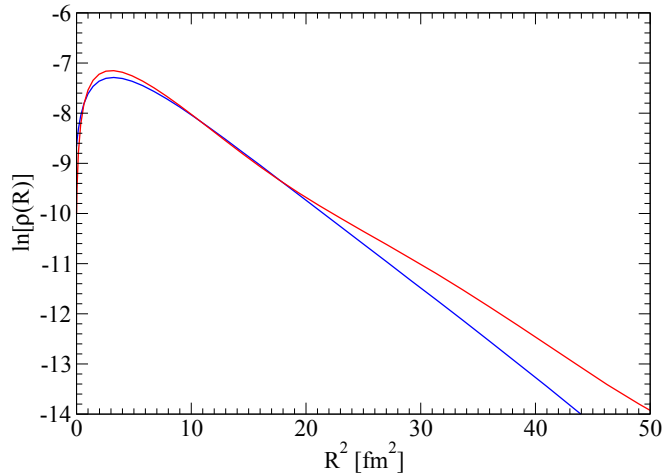}
  \caption{\label{fig:AADensity}The angular-averaged density distributions of the additional proton. The blue line is obtained by the traditional THSR wave function while the red line is obtained by the new THSR wave function(color online).}
\end{figure}

To compare our new THSR wave function with the traditional one, we show the
angular-averaged density distributions $\rho(R)$ of the valence proton in Fig.~\ref{fig:AADensity} by plotting $\ln[\rho(R)]$ as function of $R^2$. For the traditional THSR function, the Gaussian behavior of the tail is clearly observed as shown by the blue line. For the region $R < 3$ fm, the Gaussian behavior is modified because of the antisymmetrization with respect to the $^8$Be core. The new THSR wave function deviates from the traditional one for region $R > 4$ fm showing a longer tail, i.e. the density distribution spreads out compared with the Gaussian behavior. We think that the weakly bound valence proton in $^9$B is not exactly described by a Gaussian, but has a long-range density tail. The new THSR wave function may give a proof for the existence of such long-range density tails.

\begin{figure}[htbp]
  \centering
  \includegraphics[width=0.6\textwidth]{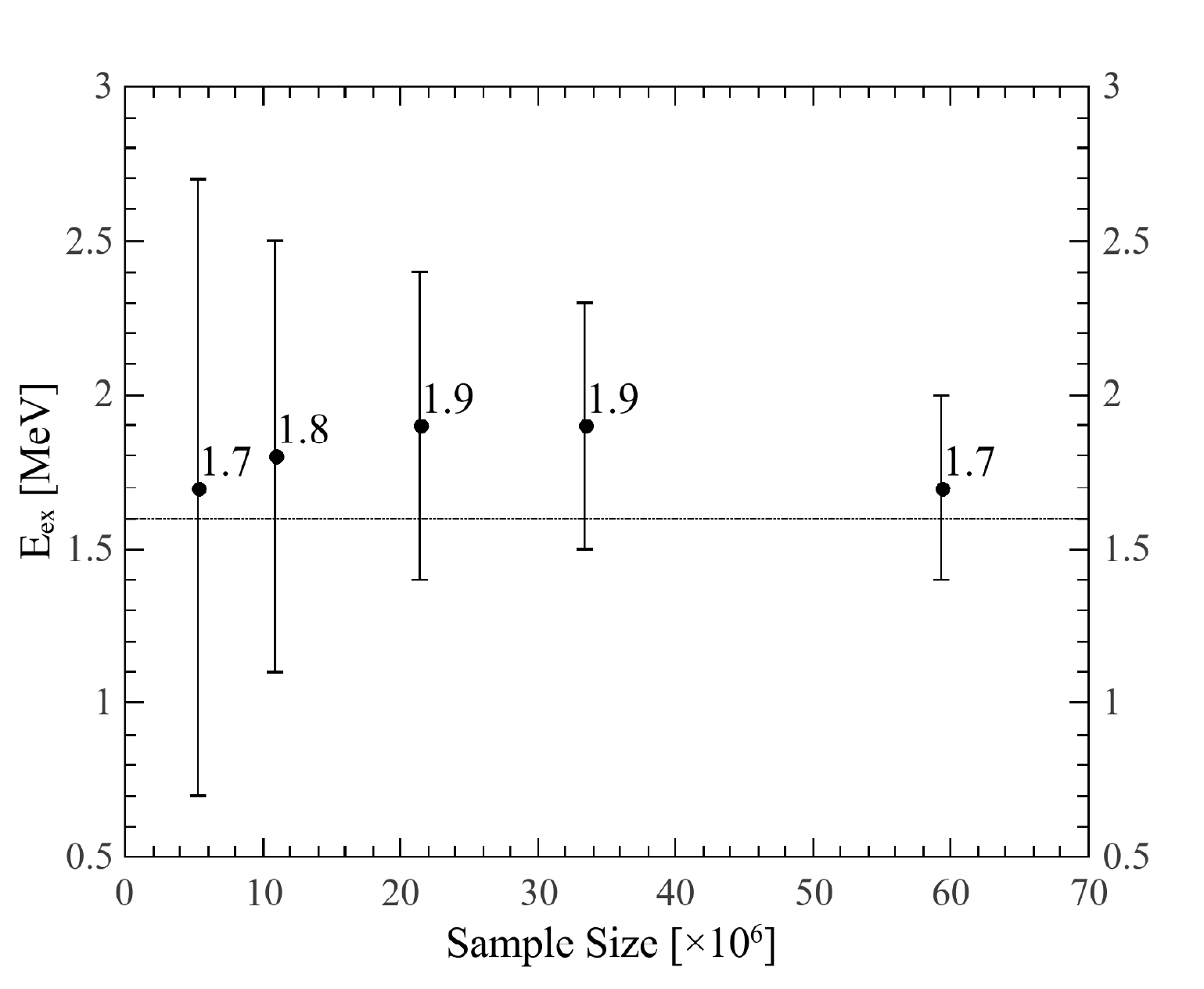}
  \caption{\label{fig:ErBarCurve}The excitation energy of the $1/2^{+}$ state in ${}^{9}$B with increasing sample size in our Monte Carlo calculation. The dotted line denotes the experimental prediction in Ref.~\cite{Tilley2004}.}
\end{figure}

\begin{figure}[htbp]
  \centering
  \includegraphics[width=0.6\textwidth]{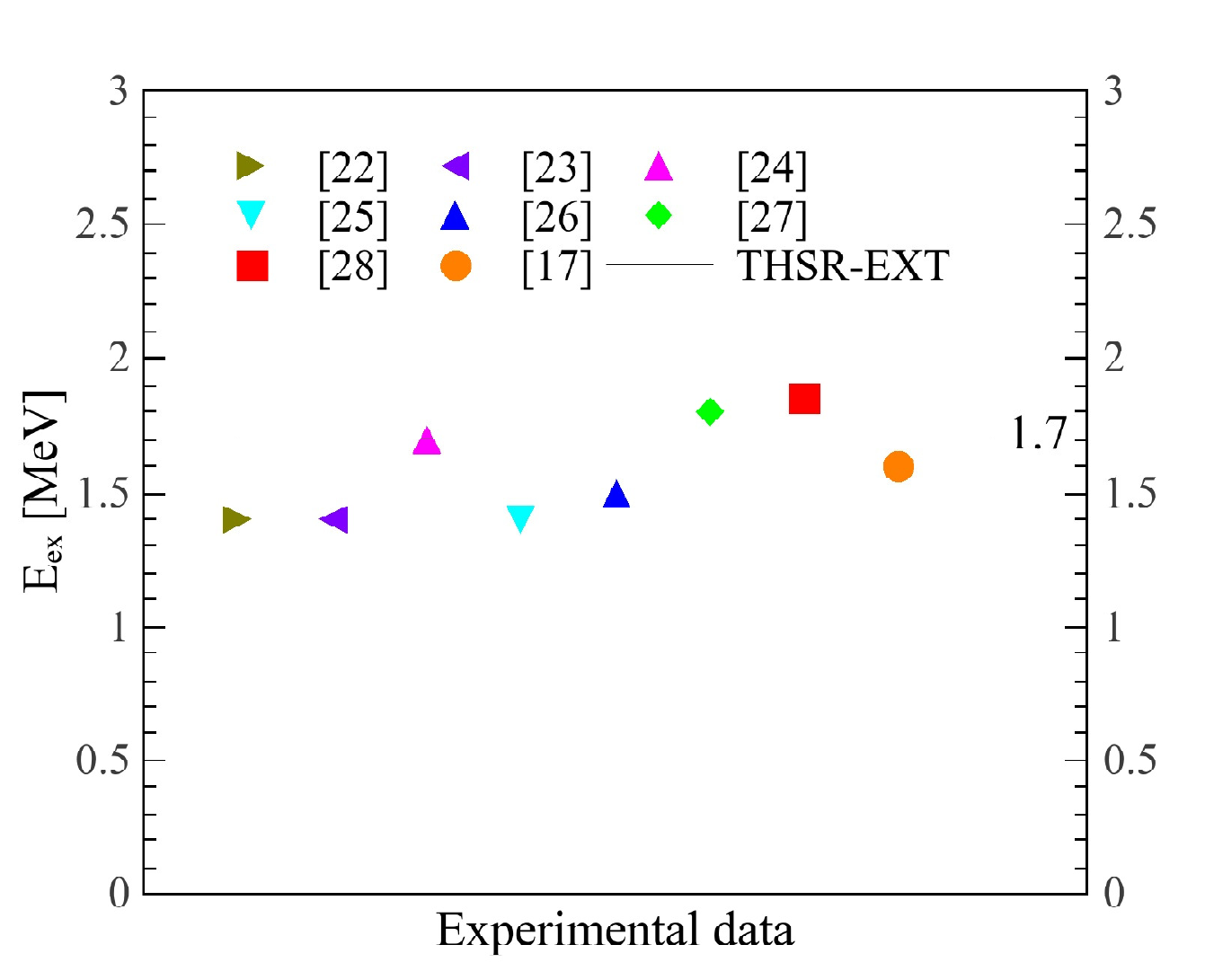}
  \caption{\label{fig:enecompare}The excited energy of the $1/2^{+}$ state of ${}^{9}$B with our new superposed THSR wave function (line) compared with many other experimental predictions (dots).}
\end{figure}

We further apply the new superposed THSR wave function to the first $1/2^+$ excited state of $^9$B. It is predicted in various theoretical models that there might be a $1/2^{+}$ excited state in ${}^9$B. However, this is still not fully confirmed by experiments except a few predictions \cite{Barker1987, Descouvemont1989, Descouvemont2001}. With our new THSR wave function, we find a local minimum by the variational calculation. Because of the limited computation power, we can not confirm this minimum with high accuracy, but we observe that this local minimum exists while we increase the precision, as shown in  Fig.~\ref{fig:ErBarCurve}. Even though this local minimum is much shallower than the error bars, our results indicate that there might be an excited state of $1/2^{+}$ in ${}^9$B with 1.7 MeV above the ground state. By comparing with many other experimental predictions as shown in Fig.~\ref{fig:enecompare}, our result of the first $1/2^{+}$ state of ${}^{9}$B fits the experiment predictions very well.


Further more, we calculate the $3/2^{+}$ state, which belongs to the rotational band of the first $1/2^{+}$ state. The energy of this state is obtained as 4.2 MeV, which is consistent with the experimental value 4.3 MeV as predicted in Ref.~\cite{Tilley2004}. This agreement further implies that the  new superposed THSR wave function provides a good description for its band head $1/2^{+}$ state.

The root-mean-square(RMS) radii are calculated for the each state of $^9$B with the new THSR wave function as shown in Table \ref{table:rms}. For the $1/2^+$ rotational band of $^9$B, the RMS radii are much larger than the ones of the other states because the decay of this state into the $0^+$ ground state of $^8$Be occurs via s-wave of the valence proton, and the decaying proton does not feel any centrifugal barrier. Hence the $1/2^+$ excited state is a resonance state which is spatially confined mainly by the Coulomb barrier. This resonance nature is the main origin of the large radii of the $1/2^+$ state and its rotational band.
\begin{table*}[htbp]
  \begin{center}
    \caption{\label{table:rms}RMS radii for each state of $^9$B obtained with the new superposed THSR wave function.}
 \begin{tabular*}{13cm}{ @{\extracolsep{\fill}} p{2cm} |p{0.5cm} p{0.5cm} p{1cm}|p{1cm}|p{1cm}p{1cm}}
    \hline
    \hline
state  &$3/2^-$  &$5/2^-$ &$7/2^-$  &$1/2^-$ &$1/2^+$ &$3/2^+$\\
	\hline
RMS(fm)    &2.81     &2.87    &3.00     &3.04    &4.05    &4.16\\
    \hline
    \hline
  \end{tabular*}
  \end{center}
\end{table*}

Further more, the computational efficiency of our variational calculation is seriously reduced by the broad distribution of nucleons. For the calculation of the $5/2^+$ state of $^9$B, the convergence can not be obtained. This is the reason why we do not include the $5/2^+$ state result in present work.

\begin{figure}[htbp]
  \centering
  \includegraphics[width=0.6\textwidth]{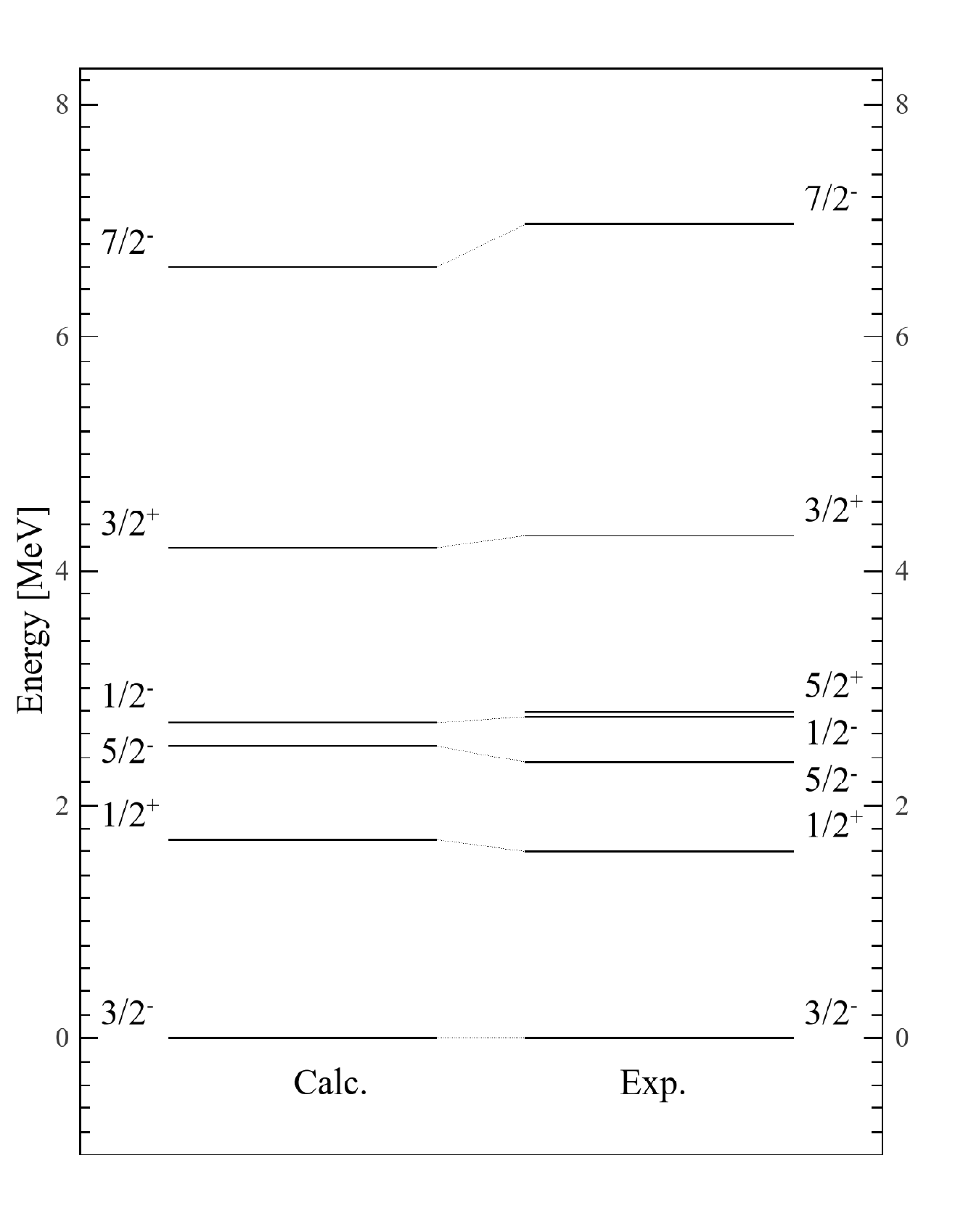}
  \caption{\label{fig:enespectrum}Theoretical and experimental energy spectra of ${}^{9}$B. The states with higher energy, which are found in experiment, are not considered in our calculation because of the limitation of computation power.}
\end{figure}
We present the total energy spectrum of ${}^{9}$B obtained from the new superposed THSR wave function in Fig.~\ref{fig:enespectrum}. We observe systematic agreements between our theoretical results and the experimental values, which shows the accuracy and efficiency of our new approach in describing general cluster states with coupling of $\alpha$-clusters and valence nucleons.

\section{Conclusion}
\label{sec:conclusion}
We formulated a new superposed THSR wave function for ${}^{9}$B with components corresponding to both the large spreading motion and the cluster-correlated motion for the valence proton. The calculated energies of the  $3/2^{-}$ rotational band is essentially improved comparing to the previous version of THSR wave function in Ref.~\cite{Lyu2015} and much better agreements with the GCM results and the experimental values is obtained. By comparing the optimized wave function with each superposed terms, it is shown clearly that both large spreading and cluster-correlated motion are essential for the description of the ${}^{9}$B nucleus. The dynamics of valence proton and the cluster-proton correlation is further discussed by density distributions of valence proton. Variational calculation results also suggest that there might be an $1/2^{+}$ excited state of $^9$B at about 1.7 MeV, which is consistent with experimental predictions. The RMS radii results of each state of $^9$B are obtained and large spatial spreading is observed for the first $1/2^+$ excited state and its corresponding rotational band because of its resonance nature. Other states are also calculated with the new superposed THSR wave function, and we produce an energy spectrum of $^9$B that agrees systematically with experimental values. This study further improves the understanding of both the physical property of $^9$B nucleus and the cluster-nucleon correlation, which is beneficial for future extension of investigations for neutron-rich or proton-rich nuclei towards the nuclear drip line.

\begin{acknowledgments}
The author would like to thank Professor Kimura and Professor Kanada-En'yo for fruitful discussions. This work is supported by the National Natural Science Foundation of China (grant nos 11535004, 11761161001, 11375086, 11120101005, 11175085, and 11235001), by the National Major State Basic Research and Development of China, grant no. 2016YFE0129300, by the Science and Technology Development Fund of Macau under grant no. 068/2011/A, and by the JSPS KAKENHI Grants No.JP16K05351.
\end{acknowledgments}

\end{document}